# Sodium Diffusion and Dynamics in Na$_2$Ti$_3$O$_7$: Neutron Scattering and Ab-initio Simulations

Ranjan Mittal[1,2*], Sajan Kumar[1,2], Mayanak K. Gupta[1$], Sanjay K. Mishra[1,2], Sanghamitra Mukhopadhyay[4], Manh Duc Le[4], Rakesh Shukla[3], Srungarpu N. Achary[2,3], Avesh K. Tyagi[2,3], and Samrath L. Chaplot[1,2]

[1]*Solid State Physics Division, Bhabha Atomic Research Centre, Mumbai, 400085, India*
[2]*Homi Bhabha National Institute, Anushaktinagar, Mumbai 400094, India*
[3]*Chemistry Division, Bhabha Atomic Research Centre, Mumbai, 400085, India*
[4]*ISIS Neutron and Muon Facility, Rutherford Appleton Laboratory, Chilton, Didcot, Oxon OX11 0QX, UK*
Email:, rmittal@barc.gov.in[*], mayankg@barc.gov.in[$]

We have performed quasielastic and inelastic neutron scattering (QENS and INS) measurements from 300 K to 1173 K to investigate the Na-diffusion and underlying host dynamics in Na$_2$Ti$_3$O$_7$. The QENS data show that the Na atoms undergo localized jumps up to 1173 K. The ab-initio molecular dynamics (AIMD) simulations supplement the measurements and show 1-d long-ranged diffusion along the *a*-axis above 1500 K. The simulations indicate that the occupancy of the interstitial site is critical for long-range diffusion. The nudged-elastic-band (NEB) calculation confirmed that the activation energy barrier is lowest for diffusion along the *a*-axis. In the experimental phonon spectra the peaks at 10 and 14 meV are dominated by Na dynamics that disappear on warming, suggesting low-energy phonons significantly contribute to large Na vibrational amplitude at elevated temperatures that enhances the Na hopping probability. We have also calculated the mode Grüneisen parameters of the phonons and thereby calculated the volume thermal expansion coefficient, which is found to be in excellent agreement with available experimental data.

## I. INTRODUCTION

Metal-ion battery research is progressively helping to make batteries lighter, faster charging, safer, stable, and more cost effective[1, 2]. Lithium-based rechargeable cells are known to have the highest density of stored electrical energy[3, 4]. However, the production of these cells may not be profitable in the long run due to the limited resources of lithium. Sodium is one of the most abundant elements existing on Earth. Recently Na-based batteries are becoming popular for the next generation. The energy density of sodium-ion batteries is relatively lower than that of lithium-ion batteries. However, they can be used for applications where energy density requirements are not stringent [5], and may be commercially viable [6-10]. The first sodium-sulfur battery was developed after the discovery of the high-temperature ionic conductor β-alumina, NaAl$_{11}$O$_{17}$ [11]. Later many Na-ion conductors have been discovered [12-15], including the sulphide compounds [16-18] that exhibit very high ionic conductivity suitable for solid electrolytes. These materials' stability, safety, and performance are related to their thermodynamic and transport behavior (such as the thermal expansion, phase transitions, electronic and ionic conductivity [19].

It is important to know the possible mechanism of diffusion and associated attributes. Such information would be helpful to design better solid-electrolytes. Further, atomic dynamics investigations provide the critical features of diffusion and thermodynamical stability of the material. Hence, it is highly desirable to have an intensive microscopic understanding of the atomic level dynamics [20, 21].

Oxides-based metal-ion conductors have attracted attention [1, 2] due to their easy synthesis, low cost, and high stability in variable temperature/pressure range. An important material, $Na_2Ti_3O_7$ has been extensively studied[22-29] for its usage as an anode material. It is known to be a very promising anode compound, especially when it is mixed with carbonaceous materials[30]. Apart from its usage as an anode in sodium batteries, the compound also finds applications in electrochemical devices [31, 32]. $Na_2Ti_3O_7$ has a layered structure (Fig. 1) [33]. The compound has a monoclinic structure[33] with the space group $P2_1/m$ (Z=2). It has zigzag stepped layers, which are composed of edge-linked distorted $TiO_6$ octahedra. The Na ions are intercalated between the layers formed by $TiO_6$ octahedra, stacked along the b-axis. The Na atoms in a layer form channels along the a-axis.

The study of ionic conductivity, thermal expansion, and structural properties of $Na_2Ti_3O_7$ has been reported [34]. The thermodynamic properties of $Na_2Ti_3O_7$ under high temperature and high pressure have been calculated[35] using DFT with local density approximation (LDA). The atomic dynamics investigation has been reported based on Raman measurements and DFT simulations [36]; these studies revealed the phonon spectra extend up to 110 meV. However, these Raman measurements are limited to the Brillouin-zone center, and theoretical studies were based on the quasiharmonic approximation, which may not be valid in the high-T regime.

The host-flexibility, vacancies and interstitial sites were proposed to be the critical factors of large-ionic conductivity in recently discovered solid electrolytes such as thiophosphates and Li- argyrodites[37-40]. Previous works on Na-based electrolytes have successfully utilized the quasielastic neutron scattering (QENS) techniques to uncover the various possible Na diffusion process in amorphous and crystalline materials [41-44]. Our interest is to investigate the possible Na diffusion and influence of host dynamics in $Na_2Ti_3O_7$ over a broad range of temperatures. Hence, we performed QENS and INS measurements covering dynamics from μeV to meV from ~300 K to ~1173 K. The QENS technique provides the characteristics of the Na diffusion process. The measurements provide essential insights about diffusion pathways and timescales of Na diffusion, which are important information to design materials for ionic diffusion. The INS data help to identify the phonon modes which show significant change with temperature. Further, we have performed AIMD and lattice dynamics simulations to supplement these measurements and to have microscopic details of the dynamical process. We have identified the critical influence of the interstitial site on the long-range Na diffusion.

## II. EXPERIMENTAL

Monoclinic α-$Na_2Ti_3O_7$ was prepared by conventional solid-state reaction of $Na_2CO_3$ and rutile-type $TiO_2$. $TiO_2$ (rutile, ≥99.9%, Alfa Aesar) was heated at 1073 K for 8h and cooled to room temperature before use. $Na_2CO_3$ (99 %, Sigma-Aldrich) was heated at 373 K for 2 h and weighed immediately after cooling to room temperature. Appropriate amounts of $Na_2CO_3$ and $TiO_2$ in the molar ratio of 2.01:3.00 for about 10 grams of $Na_2Ti_3O_7$ were mixed thoroughly using acetone as grinding media. A slight excess of $Na_2CO_3$ was desired to compensate for the loss of $Na_2O$ at higher temperatures. The homogenous mixture was filled in an alumina crucible and heated at 773 K for 12 h. The obtained powder was homogenized and pressed into pellets of about 1-inch diameter and 1 to 1.5 cm height. The pellets were heated at 1073 K for 24, and then crushed to powder and repelletized. These pellets were again heated at 1173 K for 12 h. This heating step was repeated three times to obtain single-phase monoclinic α-$Na_2Ti_3O_7$. Each time the pellets were crushed and repelletized for the next heating. The white-colored final product was characterized by Rietveld refinement of powder XRD data recorded by using Cu Kα radiation. All the observed peaks and intensities could be accounted [45] by the monoclinic $Na_2Ti_3O_7$.

Any diffusion process in the material is attributed to stochastic dynamics, which would lead to a significant broadening in the elastic signal of neutron scattering. In QENS techniques, we measure the elastic broadening at different momentum transfer $Q$ values, which is further used to determine the nature of diffusion. The measurement of QENS spectra has been done using the OSIRIS[46-49] indirect geometry time of flight spectrometer at ISIS. The QENS spectra have been measured at several temperatures from 300 K to 1173 K. In these experiments, and the sample was heated under vacuum in a thin film niobium cylindrical cell. We have also measured data from the vanadium standard for the normalization of detectors and the measurement of the energy resolution of the instrument. The instrument's resolution is 27 μeV for fixed final energy of 1.84 meV with pyrolytic graphite (002) analyzers. The analysis of data has been performed using the QENS data analysis interface as implemented Mantid [50, 51]. In case of $Na_2Ti_3O_7$, the constituent atoms scatter coherently as well as incoherently (the neutron scattering cross-sections are, Na: $\sigma_{coh}$= 1.66 barn, $\sigma_{inch}$=1.62 barn; Ti: $\sigma_{coh}$= 1.485 barn, $\sigma_{inch}$=2.87 barn; and O: $\sigma_{coh}$=4.232 barn, $\sigma_{inch}$=0.0 barn). Thus it can be noted that sodium has nearly identical coherent and incoherent cross sections[52].

The fitting of the S(Q, E) data at various Q within the dynamic range of the OSIRIS instrument is done with one Lorentzian peak and one delta function convoluted with the resolution function of the instrument. The Q-dependence of the half-width-at-half-maximum (HWHM) of the Lorentzian peak provides[53] direct information about diffusion characteristics. For long-range diffusion, HWHM in the low $Q$ regime is characterized by a $Q^2$ dependence. The Q-dependent HWHM data may be fitted using jump diffusion models, such as the Chudley-Elliott (C-E) [54] model or the H-R model [53, 55] to get the mean jump length and average jump time from one site to another site. However, for localized diffusion, HWHM is constant as a function of $Q$.

The temperature dependence of the phonon density of states in $Na_2Ti_3O_7$ was measured[56] using the MARI time of flight direct geometry spectrometer at ISIS, UK. The measurements were carried out in the energy loss mode with incident neutron energies of 180 meV and 30 meV. The two incident energies enabled to cover the entire spectral regime with a reasonable resolution of low energy phonons. The polycrystalline sample of $Na_2Ti_3O_7$ was kept in a glass tube of about 10 mm inner diameter. Measurements were carried out at several temperatures from 323 K to 1073 K. The data analysis was carried out using the Mantid [50, 51] in the incoherent one-phonon approximation. The measured scattering function $S(Q, E)$, where E and Q are the energy transfer and momentum transfer vector, respectively, is related [57-59] to the neutron-weighted phonon density of states $g^{(n)}(E)$ as follows:

$$g^{(n)}(E) = A < \frac{e^{2W(Q)}}{Q^2} \frac{E}{n(E,T) + \frac{1}{2} \pm \frac{1}{2}} S(Q,E) > \qquad (1)$$

where the + or – signs correspond to the energy loss or gain of the neutrons, respectively; $n(E,T) = [\exp(E/k_B T) - 1]^{-1}$ T is temperature, $k_B$ is the Boltzmann's constant, and $A$ is the normalization constant. The quantity between < > represents a suitable average over all $Q$ values at a given energy. $2W(Q)$ is the Debye-Waller factor. Usually the same Debye-Waller factor is assumed for all the atoms, but that assumption is not good if different atoms have very different values. In the present case, Na atom has much larger vibrational amplitude than the Ti and O atoms, so, we have omitted the Debye-Waller factor in Eq. (1) and explicitly retained it in the theoretical calculation of $g^{(n)}(E)$ as follows:

$$g^{(n)}(E) = B\sum_{k}\{\frac{\sigma_k}{m_k}\}g_k(E).\exp(-Q^2<u_k^2>/3) \qquad (2)$$

$B$ is the normalization constant, and $\sigma_k$, $m_k$, and $g_k(E)$ are, respectively, the total neutron scattering cross-section, mass, and partial density of states of the $k^{th}$ atom in the unit cell. $2W(Q)= Q^2<u_k^2>/3$, where $<u_k^2>$ is the mean-squared displacement of the atom k. The weighting factors $\frac{\sigma_k}{m_k}$ in the units of barns/amu for O, Na and Ti atoms are 0.2645, 0.1427, and 0.0909, respectively. The values of neutron scattering lengths for various atoms can be found from Ref.[52].

## III. COMPUTATIONAL DETAILS

The lattice dynamics and molecular dynamics simulations were performed within the density functional theory framework implemented in the Vienna Ab-initio Simulation Package (VASP). The calculations are performed using the projected augmented wave (PAW) formalism and with the generalized gradient approximation[60, 61] (GGA) exchange-correlation functional parameterization by Perdew, Burke, and Ernzerhof. For lattice dynamics calculations, a -k-point mesh of 8 ×4 ×4 generated using the Monkhorst--Pack method[62] is used to sample the Brillouin zone. The plane-wave kinetic energy cutoff of 1000 eV is used. The electronic self-consistency convergence criteria for electronic minimization and ionic forces were set to $10^{-8}$ eV and $10^{-3}$ eVÅ$^{-1}$, respectively.

For ab-initio molecular dynamics (AIMD) simulation, we used a 3×2×2 supercell of the monoclinic unit cell of $Na_2Ti_3O_7$. To see the effect of vacancy in the host lattice, we have introduced one Na vacancy among 48 Na sites in the 3 ×2× 2 supercell. We used an energy cutoff of 1000 eV. As we know, AIMD calculations are computationally expensive; we have taken Gamma k-point in the Brillouin zone for total energy calculations, and the energy convergence criteria for electronic minimization was $10^{-6}$ eV. The calculations are performed at 300, 1000, 1500 and 1800 K using an NVT ensemble. The temperature is controlled using a Nose-Hover thermostat[63]. The AIMD simulations were performed with 2 fs step size for ~60 ps, where the initial 5 ps data was used for equilibration, and the rest of the trajectory was utilized for the production run. The mean-squared displacement (MSD, <$u^2$>) of various atoms as a function of time is obtained from the AIMD simulations [64, 65].

The phonon calculations were performed using a supercell approach implemented in PHONOPY software[66]. A 3×2×2 supercell with 0.01 Å displacement is used to derive the force-constant matrix for subsequent phonon calculation. The supercell DFT calculations were performed with a single k-point at Gamma-point.

We have calculated the thermal expansion behavior in the quasiharmonic approximation [67]. The thermal expansion is the sum of contributions from phonon modes in the entire Brillouin zone[67-69]. In the quasiharmonic approximation, the volume thermal expansion coefficient of a crystalline material is given by the following relation:

$$\alpha_V = \frac{1}{BV} \sum_i \Gamma_i C_{Vi}(T) \qquad (4)$$

The volume dependence of the phonon frequency gives the mode Grüneisen parameter

$$\Gamma_i = -\frac{V}{E_i}\frac{dE_i}{dV} \qquad (5)$$

$C_{Vi}(T)$ is the specific heat contribution of the $i^{th}$ phonon mode (of energy $E_i$) at temperature $T$, while $B$ and $V$ are the bulk modulus and volume of the primitive unit cell, respectively.

## IV. Results and Discussion

### A. Na Diffusion, QENS Measurements, and AIMD Simulations

**Figure 2** shows the measured dynamical neutron scattering function $S(Q, E)$ of $Na_2Ti_3O_7$ summed over all the $Q$ from 300 K to 1173 K. A QENS broadening is observed at above 873 K, indicating the Na diffusion. In Fig 2(c) and Fig. S1 (Supplementary Material[70]), we have shown measured S(Q,E) at 1173 K. The $Q$-dependence of the half-width-at-half-maximum (HWHM) of fitted Lorentzian peaks at 1173 K is shown in **Fig. 2(d)**. For long-ranged diffusion, the HWHM is expected[53] to tend to zero as Q tends to zero. Surprisingly, we observed that the HWHM does not decrease at low-$Q$ regime. Further, the HWHM does not show much variation with $Q$, indicating localized Na dynamics. Hence, we estimated the time scale of this localized motion by fitting a straight line to HWHM (Fig. 3), and obtain an estimated $1/\tau \sim 0.06 \pm 0.01$ meV ($\tau \sim 11 \pm 2$ ps).

We have performed AIMD simulation to investigate the possible pathways for localized and long-range diffusion and investigate the influence of host dynamics on Na-jumps. The AIMD simulations (~50 ps) in the perfect crystalline phase do not show any signs of diffusion up to 1500 K, indicating the perfect crystalline phase is unlikely or not suitable for Na- diffusion. It is known that vacancies and defects enhance the diffusion behavior [71-73]. The calculated MSD of atoms in the structure with one vacancy in the supercell is shown in **Fig. 3(a).** We have also calculated the anisotropic components of MSD along cartesian directions (**Fig. 3(a)**) and found that Na- diffusion along a-axis (x-direction) show the fastest diffusion, while no diffusion along the b-axis (y-direction).

By observing the individual squared displacement of Na atoms (**Fig. 3(b)**) at 1000 K, we observed Na jumps by ~ 3.5 Å. At 1500 K, we find that Na atoms show jumps by ~ 3 to 4 Å and residence time of >40 ps (**Fig. 3(b)).** The residence time reduces at higher temperatures. We find that at 1000 K and 1500 K, the Na atom jumps between neighboring sites. However, the jumps are very few in our simulation up to 100 ps, indicating almost localized diffusion. At 1800 K, Na ions show several intra-channel jumps along a-axis, reflecting the long-range diffusion along the a-axis. The nature of the jumps is clearly seen in Fig 4, which shows the calculated trajectories of selected Na atoms, as well as 2-dimensional projections of the trajectories.

We may examine possible reason for the change from localized to long-ranged diffusion. Up to 1500 K, we observe isolated events that involve cooperative jumps of a few atoms to the neighboring regular atomic positions. These do not lead to long-range diffusion. We note that the probability of cooperative jumps involving simultaneous jumps of several atoms would be rather small, compared to independent individual atomic jumps. Individual jumps are not feasible in the absence of regular vacant positions or low-energy interstitial positions. At higher temperatures, when sufficient kinetic energy is available, it may be possible to observe individual jumps, where the jumping atoms halts (**Fig. 4**) for a short while at interstitial positions. Indeed, at 1800 K, such interstitial occupancy is observed, and atoms are found be undergo several jumps, leading to long-range diffusion. The calculated Na occupation probability density iso-surface plot at 1500 K (**Fig. 5**) does not show significant probability at the interstitial sites and shows localized Na diffusion, while at 1800 K, there are jumps at the interstitial positions, enabling the long-ranged diffusion.

The nudged elastic band (NEB) method is used to determine the activation energy barrier between two energy minima [74]. In the perfect crystalline case, we estimated the energy barrier by moving two Na atoms simultaneously, as shown in **Fig. 6**, finding an energy barrier of 1.0 eV. We have also computed the activation barrier for the vacancy structure. The barrier energy profile for intra-channel and inter-channel jumps are 0.27 eV and 0.5 eV, respectively (**Fig. 6**). Interestingly, we observe a much smaller barrier for intra-channel diffusion than inter-channel diffusion. An energy barrier ~ 0.34 eV is estimated from conductivity measurements (297 K to 324 K) [33], which shows a fair agreement with our calculated barrier energy in Na-vacancy. It is also important to mention that the conductivity measurements give effective barrier energy from intra- and inter-channel jump processes. To further investigate the differences in intra- and inter-channel, we have analyzed the intermediate NEB images. We found that the inter-channel Na hopping involves the rotation of neighboring $TiO_6$, while an intra-channel jump has the least effect on neighboring $TiO_6$. We note that a relatively firm 2-d layered structure costs significantly to rotate the $TiO_6$ units and, in turn, leads to larger barrier energy.

The barrier energy for the vacancy structure between two nearest Na-sites along the *a*-axis has also been calculated from the free-energy landscape of the Na ion ($F_{Na}(r)$) at 1800 K using the following relation[75]:

$$F_{Na}(r) = -k_B T \ln (P_{Na}(r))$$

Here $k_B$, $T$, and $P_{Na}(r)$ are the Boltzmann constant, temperature, and probability of Na occupancy at a distance $r$ from the equilibrium Na position. The calculated $F_{Na}(r)$ is shown in **Fig 5(b)**, which indicates the barrier of $E_a$ ~ 0.60 eV for Na migration. On the other hand, the nudged elastic band (NEB) method gives (**Fig 6(a)**) a value of 0.27 eV for the intra-channel diffusion. It may be noted that in NEB method, we simulated a single Na jump from a regular site to a vacant site, while $F_{Na}(r)$ is estimated from AIMD trajectory and includes all possible events such as single and correlated jumps; hence it may give a more realistic estimate of $E_a$. The energy barrier calculations from both the methods show a local minimum in the barier energy profiles at half of the distance (~1.9 Å) between two Na atoms, indicating interstitial positions along the *a*-axis that may enable the long-range diffusion.

Soft bond valence sum analysis[76] based on structure data have been successfully used to explore the diffusion pathways in many systems as an initial guess and sometime predict very close pathways [77, 78].

In order to search other possibility and supplement the AIMD predictions, we have performed soft bond valence sum analysis using the structural data from Ref [33]. The contribution of individual Na sites to the Na-ion conduction has been estimated by performing bond valence energy landscape (BVEL) map analysis implemented in FULLPROF software[79] (**Fig. 7**). For the BVEL map, we have chosen an iso-energy surface of 0.34 eV, which corresponds to experimentally estimated barrier energy[33] from conductivity measurements. The analysis reveals (**Fig. 7**) that Na ions are mobile in the a-c plane. As noted above, due to the layered structure of the material, the conductivity along b-axis is fully suppressed. The BVEL assumed equal bond valance energy for both the intra- and inter-channel bonds, and this might be the reason for two-dimensional diffusion in the a-c plane, in contrast to AIMD and NEB predictions of one-dimensional diffusion. Further, BVEL analysis did not predict other new pathways for Na diffusion, which confirms that our AIMD simulations (~50 ps) have taken care of all possible pathways.

## B. INS Measurements and Lattice Dynamics

The measured temperature dependence of the phonon spectra of $Na_2Ti_3O_7$ from 323 K to 1023 K are shown in **Fig. 8**. The measurements performed with incident energy of 180 meV, provides an energy resolution ~7 meV, enable the collection of data in the entire spectral range. The large incident energy has enabled to collect data at high momentum transfer vector Q range from 4 to 16 Å$^{-1}$. However, the data collection at such large Q values would be contaminated from large multi-phonon contribution, which has been corrected using the standard formalism as discussed in Ref [80].

The phonon spectra measured with incident energy of 180 meV show sharp peaks at 30 meV, 38 meV, 46 meV, 50 meV, 60 meV, 80 meV, 90 meV and 107 meV. The peak structure is gradually broadened on warming, and peaks in the DOS start merging at 573 K. To identify the individual atomic contribution in the INS spectra, we calculated the atomic-species resolved neutron-weighted DOS and compared it with the 323 K measurement (**Fig 9(a)**). The calculation shows (**Fig 9(a)**) that the measured neutron-weighted DOS is dominated by O-atom dynamics. **Figure 9 (c)** shows the total and partial density of states. The low energy spectra below 20 meV have a significant contribution from Na dynamics. Ti has a low- to mid-energy range contribution. The calculated range of phonon spectra matches very well with our experimental data and Raman measurements [36]. It is interesting to note that the Na dominant contribution at low energy reflects the softer bonding of Na to the host. However, the significant overlap of Na-DOS with O- and Ti-DOS also indicates that the Na dynamics is affected by the dynamics of the host elements.

The INS spectra have also been measured with incident neutron energy of 30 meV to probe the low energy phonons with a better energy resolution of ~1 meV. The measurements enabled the collection of neutron spectra up to 22 meV in the Q range from 2 to 7 Å$^{-1}$. The low energy spectra show peaks at 10 and 14 meV due to Na atoms, which get weakened and broadened with the increase in temperature (**Fig. 8**). The calculated neutron-weighted phonon density of states at 300 and 1100 K is shown in **Fig. 9(b).** We found that the contribution due to the Na atom to the phonon spectrum at 10 and 14 meV decreases substantially on increase of temperature from 300 K to 1100 K due to the increase in the Debye-Waller factor. Thus, the disappearance of the 10 and 14 meV peaks on warming (**Fig. 7**) is attributed to the large vibrational amplitude of Na at high temperatures. The increase in the vibrational amplitude of the Na atoms at high temperature may enhance the Na hoping probability. Hence, these low-energy Na modes could play an important role in Na diffusion.

## C. Calculated and Experimental Thermal Expansion Behavior

We have calculated the pressure dependence of the phonon spectrum of Na$_2$Ti$_3$O$_7$ and obtained the mode Grüneisen parameters in the entire Brillouin zone. The calculated mode Grüneisen parameters Γ of phonons of energy E, averaged over the entire Brillouin zone, as a function of phonon energy (E) is shown in **Fig. 10(a)**. It can be seen that the low-energy phonon modes below 20 meV have very large positive values of Γ up to +10. The calculated volume thermal expansion coefficient ($\alpha_V$) is shown in inset of **Fig. 10(a)**. Our calculations are found to be in excellent agreement (**Fig. 10(b)**) with the available experimental data from literature [34]. Previously, $\alpha_V$ for Na$_2$Ti$_3$O$_7$ has been calculated [35] within local density approximation (LDA) and found to be is ~21 × 10$^{-6}$ K$^{-1}$ at 1000 K, which is significantly underestimated as compared to our calculations of 44 × 10$^{-6}$ K$^{-1}$ at 1000 K within generalized gradient approximation (GGA) as well as the experimental value (inset of **Fig. 10(a)**).

Further, we have calculated the contribution (**Fig. 10(c)**) to the volume thermal expansion coefficient from phonons of energy E, averaged over the entire Brillouin zone, as a function of phonon energy (E) at 300 K. We find that the phonon modes of up to 40 meV contribute the most to $\alpha_V$. The contribution of phonons of energy E to the mean-squared amplitude (u$^2$) of various atoms at 300 K shows (**Fig. 10(d)**) that Na has very large values for phonons of energy below 20 meV, while contributions from Ti and O is very small. The structure of Na$_2$Ti$_3$O$_7$ consists of layers of distorted TiO$_6$ octahedra, with Na ions intercalated in between the layers. The large value of u$^2$ for phonon modes below 20 meV indicates that large displacement of Na within a layer is mainly responsible for large values of $\alpha_V$ for Na$_2$Ti$_3$O$_7$.

## V. CONCLUSIONS

In this paper, based on the QENS experiments and AIMD simulations, the Na diffusion in $Na_2Ti_3O_7$ has been found to be one-dimensional through channels along the crystallographic a-axis. The diffusion at low temperatures is found to be localized, while above 1500 K atoms halt at interstitial positions and undergo several jumps, leading to long-range diffusion. The calculated activation energy barrier for the intra-channel diffusion (along the a-axis) is found to be much smaller in comparison to that for the inter-channel diffusion, which is consistent with the AIMD simulations. Further, we have reported INS measurements of the phonon DOS and lattice dynamics calculations. The Na phonon modes at low energies below 20 meV are found to weaken significantly in intensity at high temperatures due to the large Na vibrational amplitudes. The volume thermal expansion coefficient has also been calculated using the quasiharmonic approximation. The low-energy Na phonon modes are found to lead to a high value of the volume thermal expansion coefficient in $Na_2Ti_3O_7$.

## ACKNOWLEDGEMENT

The use of ANUPAM super-computing facility at BARC is acknowledged. Authors thank the Department of Science and Technology, India (SR/NM/Z-07/2015) for the financial support and Jawaharlal Nehru Centre for Advanced Scientific Research (JNCASR) for managing the project. The authors also thank STFC, UK, for the beam-time at ISIS and also availing the travel support from the Newton fund, ISIS, UK. Experiments at the ISIS Neutron and Muon Source were supported by a beamtime allocation RB1868015 and RB1910264 from the Science and Technology Facilities Council. SLC thanks the Indian National Science Academy for the financial support of the INSA Senior Scientist position.

1. J. W. Choi and D. Aurbach, *Nature Reviews Materials*, 2016, **1**, 16013.
2. Y. Kato, S. Hori, T. Saito, K. Suzuki, M. Hirayama, A. Mitsui, M. Yonemura, H. Iba and R. Kanno, *Nature Energy*, 2016, **1**, 16030.
3. M. Wakihara, *Materials Science Engineering: R: Reports*, 2001, **33**, 109-134.
4. J. B. Goodenough, *Energy Storage Materials*, 2015, **1**, 158-161.
5. B. L. Ellis and L. F. Nazar, *Current Opinion in Solid State Materials Science*, 2012, **16**, 168-177.
6. Y.-E. Zhu, X. Qi, X. Chen, X. Zhou, X. Zhang, J. Wei, Y. Hu and Z. Zhou, *Journal of Materials Chemistry A*, 2016, **4**, 11103-11109.
7. X. Li, S. Guo, F. Qiu, L. Wang, M. Ishida and H. Zhou, *Journal of Materials Chemistry A*, 2019, **7**, 4395-4399.
8. J. Alvarado, G. Barim, C. D. Quilty, E. Yi, K. J. Takeuchi, E. S. Takeuchi, A. C. Marschilok and M. M. Doeff, *Journal of Materials Chemistry A*, 2020, **8**, 19917-19926.
9. G. Yan, S. Mariyappan, G. Rousse, Q. Jacquet, M. Deschamps, R. David, B. Mirvaux, J. W. Freeland and J.-M. Tarascon, *Nature Communications*, 2019, **10**, 1-12.
10. J. Wang, Y. Wang, D. H. Seo, T. Shi, S. Chen, Y. Tian, H. Kim and G. Ceder, *Advanced Energy Materials*, 2020, **10**, 1903968.
11. F. Yung, Y. Yao and J. T. Kummer, *Journal of Inorganic Nuclear Chemistry*, 1967, **29**, 2453-2475.
12. S.-H. Bo, Y. Wang and G. Ceder, *Journal of Materials Chemistry A*, 2016, **4**, 9044-9053.
13. C. Yu, S. Ganapathy, N. J. J. de Klerk, E. R. H. van Eck and M. Wagemaker, *Journal of Materials Chemistry A*, 2016, **4**, 15095-15105.
14. S. Muy, J. C. Bachman, L. Giordano, H.-H. Chang, D. L. Abernathy, D. Bansal, O. Delaire, S. Hori, R. Kanno, F. Maglia, S. Lupart, P. Lamp and Y. Shao-Horn, *Energy & Environmental Science*, 2018, **11**, 850-859.
15. T. Krauskopf, S. Muy, S. P. Culver, S. Ohno, O. Delaire, Y. Shao-Horn and W. G. Zeier, *Journal of the American Chemical Society*, 2018, **140**, 14464-14473.
16. A. Hayashi, K. Noi, A. Sakuda and M. Tatsumisago, *Nature Communications*, 2012, **3**, 856.
17. K. Noi, A. Hayashi and M. Tatsumisago, *Journal of Power Sources*, 2014, **269**, 260-265.
18. M. Guin and F. Tietz, *Journal of Power Sources*, 2015, **273**, 1056-1064.
19. P. H. Jampani, O. Velikokhatnyi, K. Kadakia, D. H. Hong, S. S. Damle, J. A. Poston, A. Manivannan and P. N. Kumta, *Journal of Materials Chemistry A*, 2015, **3**, 8413-8432.
20. Y. Wang, W. D. Richards, S. P. Ong, L. J. Miara, J. C. Kim, Y. Mo and G. Ceder, *Nature Materials*, 2015, **14**, 1026.
21. Q. Wang, J. A. Jackson, Q. Ge, J. B. Hopkins, C. M. Spadaccini and N. X. Fang, *Physical Review Letters*, 2016, **117**, 175901.
22. L. Luo, Y. Zhen, Y. Lu, K. Zhou, J. Huang, Z. Huang, S. Mathur and Z. Hong, *Nanoscale*, 2020, **12**, 230-238.
23. P. Senguttuvan, G. Rousse, V. Seznec, J.-M. Tarascon and M. R. Palacin, *Chemistry of Materials*, 2011, **23**, 4109-4111.
24. Y. Cao, Q. Ye, F. Wang, X. Fan, L. Hu, F. Wang, T. Zhai and H. Li, *Advanced Functional Materials*, 2020, 2003733.
25. T. K. Saothayanun, T. T. Sirinakorn and M. Ogawa, *Inorganic Chemistry*, 2020, **59**, 4024-4029.
26. M. A. Tsiamtsouri, P. K. Allan, A. J. Pell, J. M. Stratford, G. Kim, R. N. Kerber, P. C. Magusin, D. A. Jefferson and C. P. Grey, *Chemistry of Materials*, 2018, **30**, 1505-1516.
27. M. Mori, Y. Kumagai, K. Matsunaga and I. Tanaka, *Physical Review B*, 2009, **79**, 144117.
28. L. Zhang, L. Chen, X. Fan, X. Xiao, J. Zheng and X. Huang, *Journal of Materials Chemistry A*, 2017, **5**, 6178-6185.
29. C. Wu, W. Hua, Z. Zhang, B. Zhong, Z. Yang, G. Feng, W. Xiang, Z. Wu and X. Guo, *Advanced Science*, 2018, **5**, 1800519.
30. Z. Zhou, H. Xiao, F. Zhang, X. Zhang and Y. Tang, *Electrochimica Acta*, 2016, **211**, 430-436.

31. Y. Wei, L. Shen, Z. Wang, W.-D. Yang, H. Zhu and H. Liu, *International Journal of Hydrogen Energy*, 2011, **36**, 5088-5095.
32. M. Holzinger, J. Maier and W. Sitte, *Solid State Ionics*, 1996, **86**, 1055-1062.
33. Y. Fukuzumi, W. Kobayashi and Y. Moritomo, *Journal of Advances in Nanomaterials*, 2016, **1**, 39.
34. M. Dynarowska, J. Kotwiński, M. Leszczynska, M. Marzantowicz and F. Krok, *Solid State Ionics*, 2017, **301**, 35-42.
35. H. Zhang and H. Wu, *Physica Status Solidi. A, Applied Research*, 2008, **245**, 37-43.
36. F. L. R. Silva, *J. Raman Spectrosc*, 2018, **49**, 538-548.
37. L. Zhou, A. Assoud, Q. Zhang, X. Wu and L. F. Nazar, *Journal of the American Chemical Society*, 2019, **141**, 19002-19013.
38. B. J. Morgan, *Chemistry of Materials*, 2021, **33**, 2004-2018.
39. R. Schlenker, A.-L. Hansen, A. Senyshyn, T. Zinkevich, M. Knapp, T. Hupfer, H. Ehrenberg and S. Indris, *Chemistry of Materials*, 2020, **32**, 8420-8430.
40. Z. Zhu, I.-H. Chu, Z. Deng and S. P. Ong, *Chemistry of Materials*, 2015, **27**, 8318-8325.
41. T. Willis, D. Porter, D. Voneshen, S. Uthayakumar, F. Demmel, M. Gutmann, M. Roger, K. Refson and J. Goff, *Scientific reports*, 2018, **8**, 3210.
42. M. K. Gupta, S. K. Mishra, R. Mittal, B. Singh, P. Goel, S. Mukhopadhyay, R. Shukla, S. N. Achary, A. K. Tyagi and S. L. Chaplot, *Physical Review Materials*, 2020, **4**, 045802.
43. Q. Chen, N. H. Jalarvo and W. Lai, *Journal of Materials Chemistry A*, 2020, **8**, 25290-25297.
44. Q. Zhang, C. Zhang, Z. D. Hood, M. Chi, C. Liang, N. H. Jalarvo, M. Yu and H. Wang, *Chemistry of Materials*, 2020, **32**, 2264-2271.
45. O. Yakubovich and V. Kireev, *Crystallography Reports*, 2003, **48**, 24-28.
46. R. Mittal and etal, *Na-Dynamics in Solid Ionic Conductors: 11Al2O3-Na2O and Na2Ti3O7, STFC ISIS Neutron and Muon Source*, https://doi.org/10.5286/ISIS.E.RB1910264
47. F. Demmel, D. McPhail, J. Crawford, K. Pokhilchuk, V. G. Sakai, S. Mukhopadhyay, M. T. F. Telling, F. J. Bermejo, N. T. Skippe and F. Fernandez-Alonso, *Eur. Phys. J. Web Conf.*, 2015, **83**, 03003.
48. M. T. Telling, S. I. Campbell, D. Engberg, D. M. y Marero and K. H. Andersen, *Physical Chemistry Chemical Physics*, 2016, **18**, 8243-8243.
49. B. L. Vidal, E. Oram, R. A. Baños, L. C. Pardo, S. Mukhopadhyay and F. Fernandez-Alonso, *Journal of Physics: Conference Series*, 2018, **1021**, 012012.
50. S. Mukhopadhyay, B. Hewer, S. Howells and A. Markvardsen, *Physica B: Condensed Matter*, 2019.
51. O. Arnold, J.-C. Bilheux, J. Borreguero, A. Buts, S. I. Campbell, L. Chapon, M. Doucet, N. Draper, R. F. Leal and M. Gigg, *Nuclear Instruments and Methods in Physics Research Section A: Accelerators, Spectrometers, Detectors and Associated Equipment*, 2014, **764**, 156-166.
52. V. F. Sears, *Neutron News*, 1992, **3**, 26-37.
53. M. Bee, *Quasi Elastic Neutron Scattering; Adam Hilger*, IOP Publishing Ltd: Bristol, England, 1988.
54. C. T. Chudley and R. J. Elliott, *Proceedings of the Physical Society*, 1961, **77**, 353-361.
55. P. L. Hall and D. Ross, *Molecular Physics*, 1981, **42**, 673-682.
56. S. L. Chaplot and etal, *Na-Dynamics in Solid Ionic Conductors: 11Al2O3-Na2O and Na2Ti3O7, STFC ISIS Neutron and Muon Source*, https://doi.org/10.5286/ISIS.E.RB1868015
57. K. S. D. L. Price *Neutron scattering*, Academic Press, Orlando, 1986.
58. J. M. Carpenter and D. L. Price, *Physical Review Letters*, 1985, **54**, 441-443.
59. S. Rols, H. Jobic and H. Schober, *Comptes Rendus Physique*, 2007, **8**, 777-788.
60. K. Burke, J. Perdew and M. Ernzerhof, *Phys. Rev. Lett*, 1997, **78**, 1396.
61. J. P. Perdew, K. Burke and M. Ernzerhof, *Physical Review Letters*, 1996, **77**, 3865.
62. H. J. Monkhorst and J. D. Pack, *Physical Review B*, 1976, **13**, 5188-5192.
63. S. Nosé, *The Journal of Chemical Physics*, 1984, **81**, 511-519.
64. M. P. Allen and D. J. Tildesley, *Computer simulation of liquids*, Oxford university press, 2017.
65. A. K. Sagotra, D. Chu and C. Cazorla, *Physical Review Materials*, 2019, **3**, 035405.

66. A. Togo and I. Tanaka, *Scripta Materialia*, 2015, **108**, 1-5.
67. R. Mittal, M. K. Gupta and S. L. Chaplot, *Progress in Materials Science*, 2018, **92**, 360-445.
68. R. Mittal and S. L. Chaplot, *Physical Review B*, 1999, **60**, 7234-7237.
69. R. Mittal, S. L. Chaplot, H. Schober and T. A. Mary, *Physical Review Letters*, 2001, **86**, 4692-4695.
70. *See Supplementary Material for experimental QENS spectra at 1173 K.* .
71. J. Sugiyama, K. Mukai, Y. Ikedo, H. Nozaki, M. Månsson and I. Watanabe, *Physical Review Letters*, 2009, **103**, 147601.
72. T. Shibata, Y. Fukuzumi, W. Kobayashi and Y. Moritomo, *Scientific reports*, 2015, **5**, 1-4.
73. E. Rucavado, F. Landucci, M. Döbeli, Q. Jeangros, M. Boccard, A. Hessler-Wyser, C. Ballif and M. Morales-Masis, *Physical Review Materials*, 2019, **3**, 084608.
74. G. Henkelman and H. J. T. J. o. c. p. Jónsson, 2000, **113**, 9978-9985.
75. J. Wang, J. Ding, O. Delaire and G. Arya, *ACS Applied Energy Materials*, 2021, DOI: 10.1021/acsaem.1c01237.
76. S. Adams, *Acta Crystallographica Section B: Structural Science*, 2001, **57**, 278-287.
77. H. Chen, L. L. Wong and S. Adams, *Acta Crystallographica Section B: Structural Science, Crystal Engineering and Materials*, 2019, **75**, 18-33.
78. D. Urushihara, S. Kawaguchi, R. Harada, T. Asaka and K. Fukuda, *Materials Letters*, 2020, **277**, 128373.
79. J. Rodríguez-Carvajal, *Physica B: Condensed Matter*, 1993, **192**, 55-69.
80. B. Fultz, T. Kelley, J. Lin, J. Lee, O. Delaire, M. Kresch, M. McKerns and M. Aivazis, *Experimental inelastic neutron scattering: Introduction to DANSE*, 2009.

FIG 1 (Color online) Crystal structure of $Na_2Ti_3O_7$. The color scheme is Ti: Blue, Na: Yellow and O: Red.

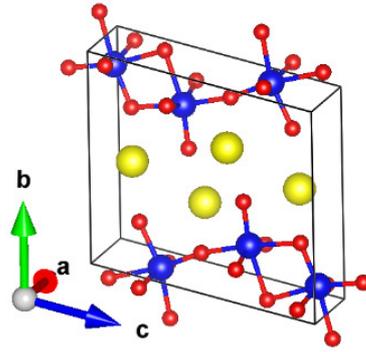

FIG 2 (Color online). (a) Observed dynamical neutron scattering function S(Q, E) of $Na_2Ti_3O_7$ integrated over all wave-vector transfers (Q) at various temperatures. (b) Comparison of as observed dynamical neutron scattering function S(Q, E) of $Na_2Ti_3O_7$ integrated over all Q at 300 K and 1173 K and the instrumental resolution (vanadium at 300 K). The data have been normalized to unity at the peak position. (c) Fit of one Lorentzian peak and one Gaussian function convoluted with the resolution function at a selected Q= 1.30 Å$^{-1}$ slice of S(Q,E), and a linear background. (d) The Q dependent of the variation of the amplitude of half width at half maximum (HWHM) of the Lorentzian peak extracted from the dynamic neutron scattering function S(Q, E) of $Na_2Ti_3O_7$ at 1173 K. The fit of the experimental data to the localized diffusion model (red line) model is also shown.

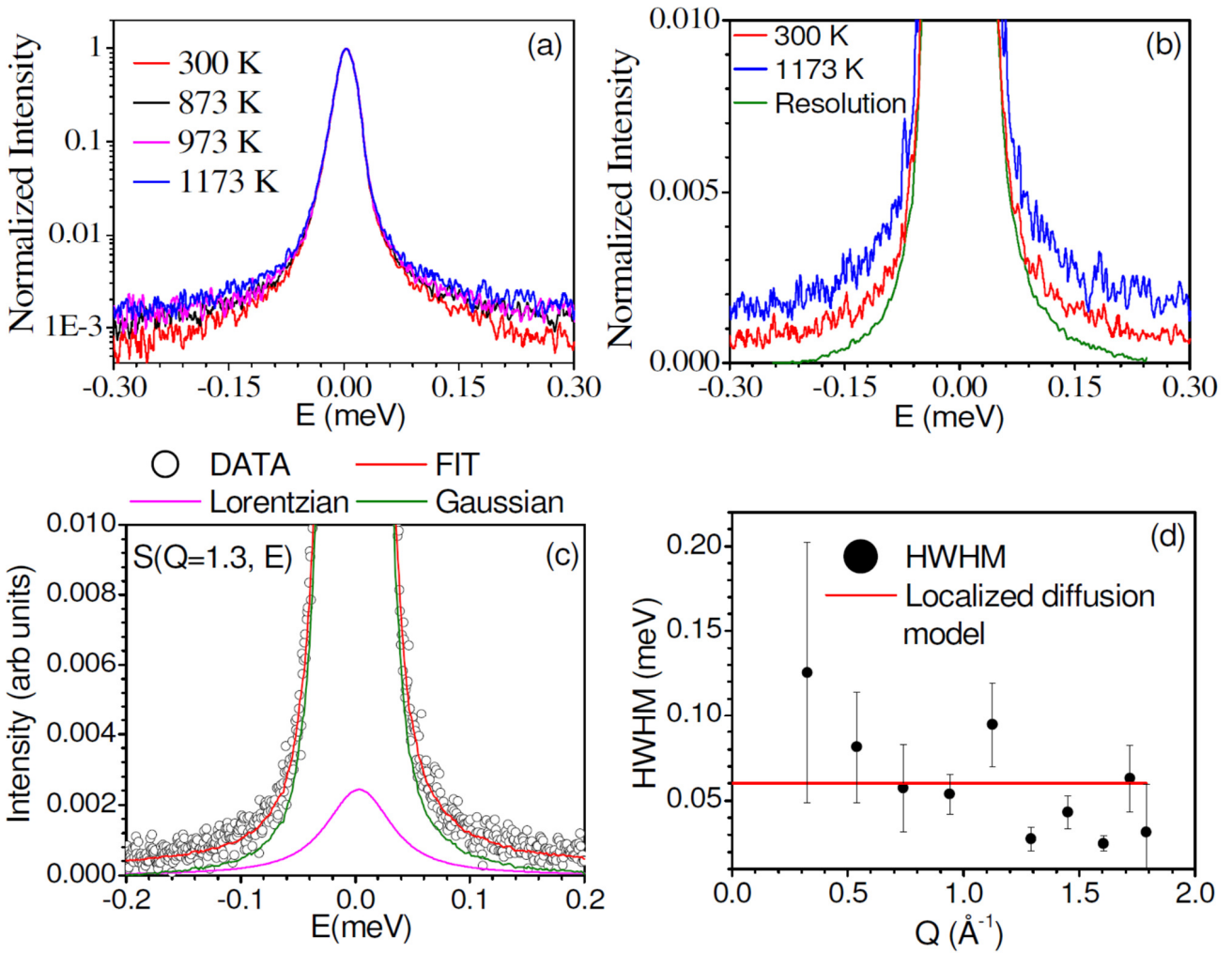

FIG 3 (Color online). (a) Time-dependent mean-squared displacements (MSD; $<u^2> = <u_x^2> + <u_y^2> + <u_z^2>$, for various atoms in $Na_2Ti_3O_7$ with one Na vacancy in a (3×2×2) supercell. Inset in (a) shows the calculated Na MSD components along x, y, and z-axis. (b) The calculated $u^2$ of individual Na atoms is represented by different colors. The 3-d trajectories of selected Na atoms (marked by thick red line) are shown in Fig 4.

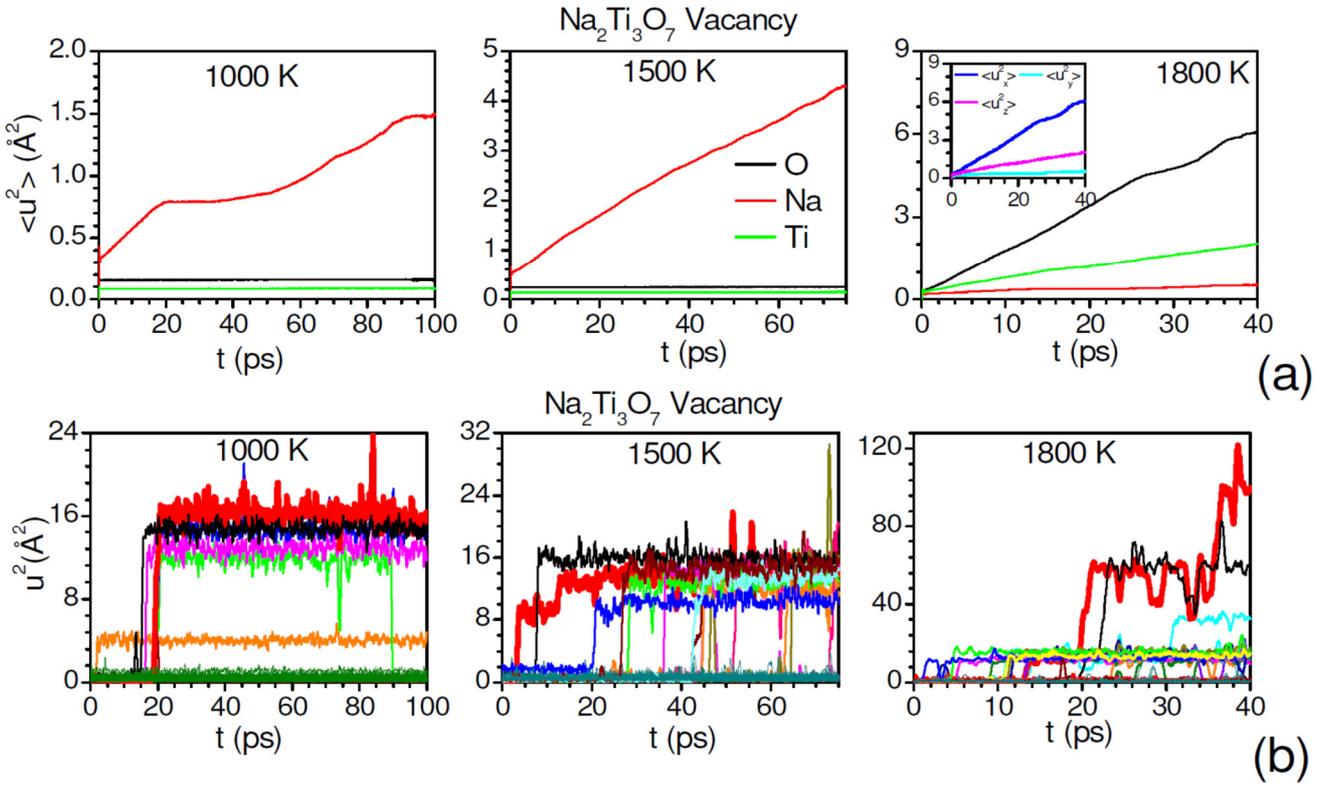

FIG 4 (Color online) Trajectory of selected diffusing Na atoms in $Na_2Ti_3O_7$ at 1000, 1500 and 1800 K. Red, yellow and blue spheres represent O, Na and Ti atoms respectively at their lattice sites. The time-dependent positions of the selected Na atoms are shown by green color dots. The numbers below each frame indicate the temperature of the simulation and duration of trajectory of Na. For clarity we have shown zoom of a part of the 3×2×2 supercell to highlight the path of Na atom. The $u^2$ of the selected Na atoms in **Fig. 3** is shown with thick solid lines. The 2-dimensional projections of the trajectory in the X-Z and X-Y planes are also shown for each of the trajectory. The fractional co-ordinates (x, y, z) shown with respect to the 3×2×2 supercell.

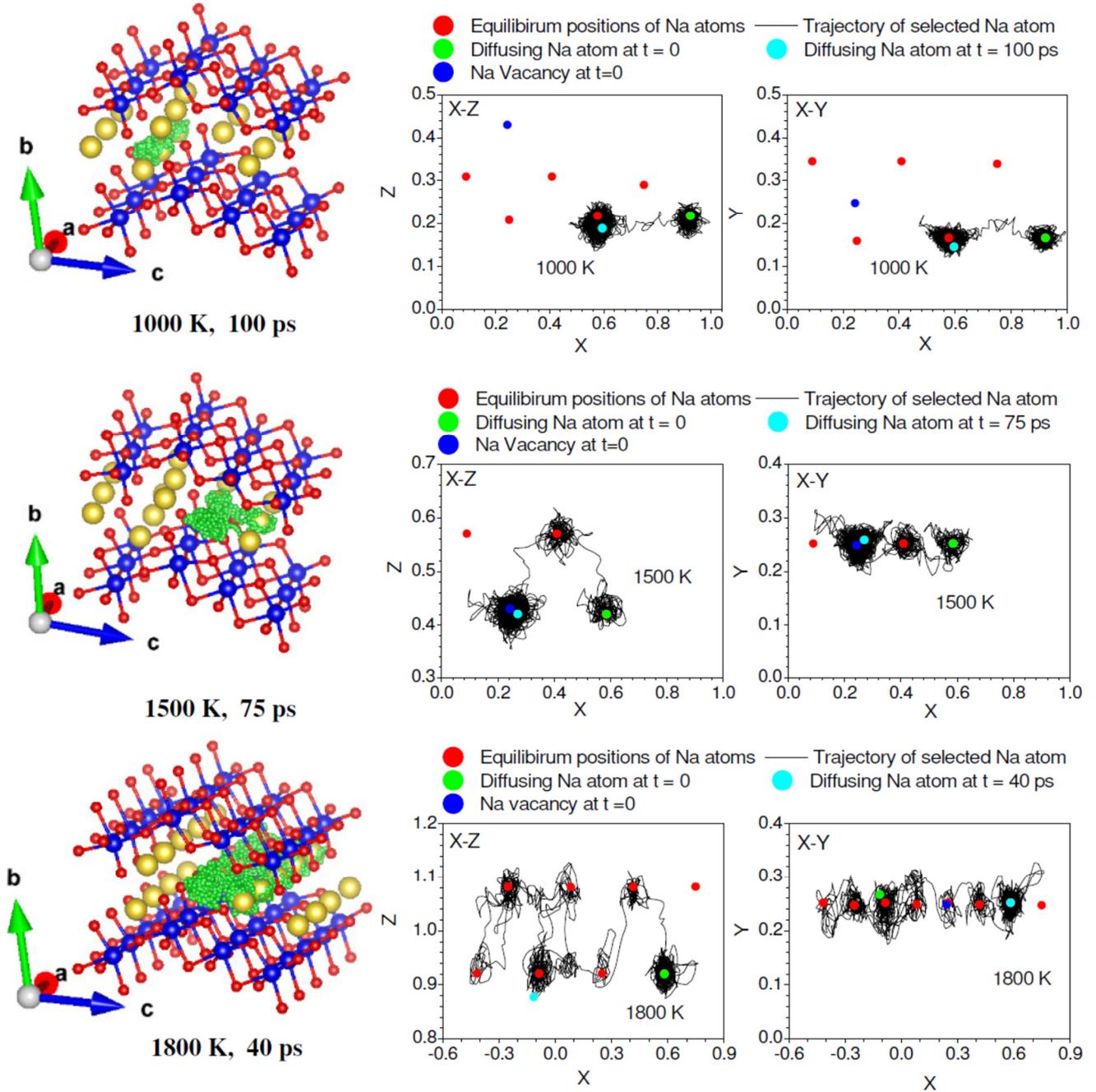

FIG 5. (Color online) **(a)** The calculated Na iso-surface probability (green dots) in the vacancy structure of $Na_2Ti_3O_7$ at 1500 K and 1800 K. Yellow spheres show equilibrium Na sites. Both the panels are shown in the a-c plane, where the c-axis is verticle and the a-axis is horizontal. **(b)** The calculated free energy ($F_{Na}(r)$) at 1800 K from Na probability distribution between two nearest Na-sites along the *a*-axis. 'r' is the distance from Na sites along *a*-axis.

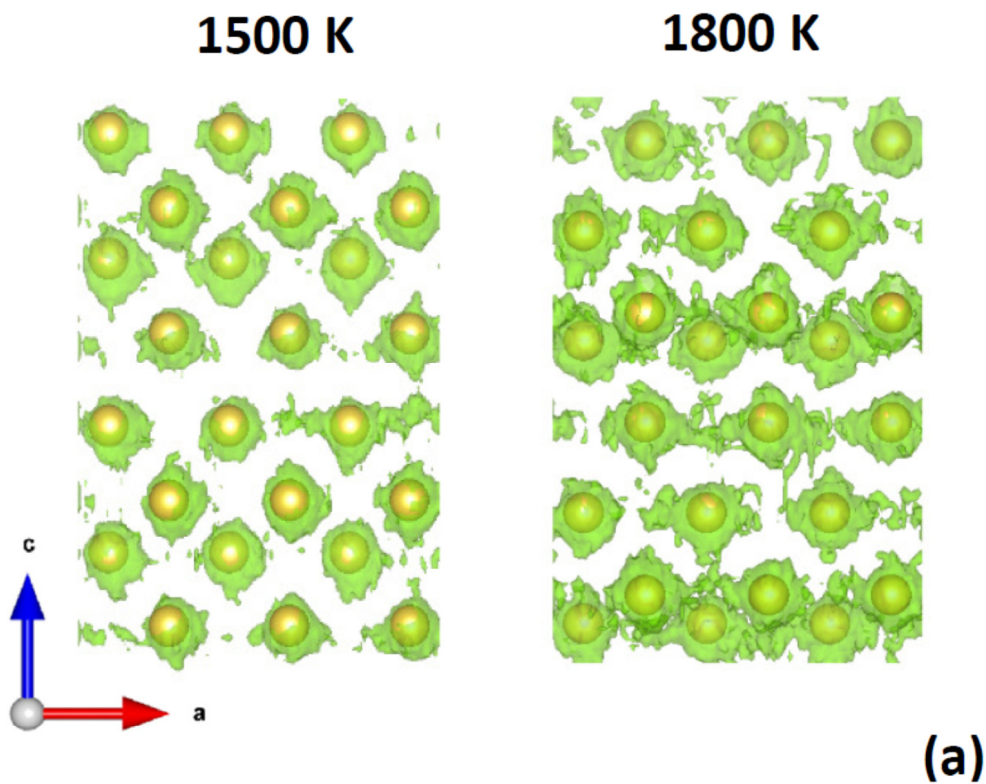

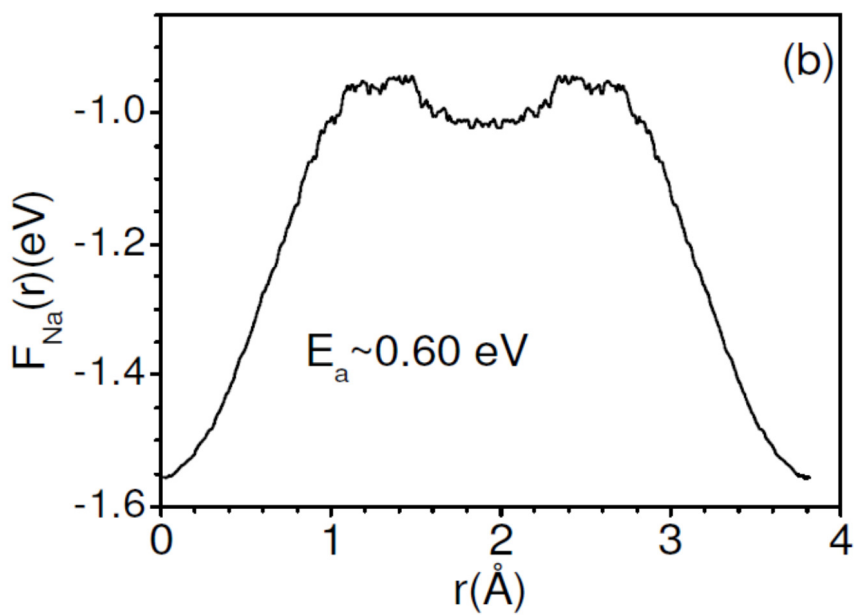

FIG 6 (Color online) (a) The estimated activation energy barrier from nudged elastic band (NEB) method. (b) The path of Na atom for estimation of energy barrier in perfect crystalline structure. The correlated motion of Na1 and Na2 are shown by brown and orange circles. Na1 moves to the position of Na2, while Na2 moves to an interstitial position. (c) and (d) are the paths of Na atom for estimation of energy barrier in the vacancy structure, for Na movement in the intra-channel and inter-channel cases, respectively. In (c) and (d) we have shown zoom of a part of the 3×2×2 supercell to highlight the path of Na atom for estimation of energy barrier. The color scheme is Ti: blue and O: red. Na at regular sites is shown by yellow color. The green color in (c) and (d) is Na vacancy, and the Na-path connects the green circle with a yellow circle. Small brown and orange circles are the position of Na atoms in various images of NEB simulation. Magenta circle in (b) is the position of Na2 atom at interstitial site.

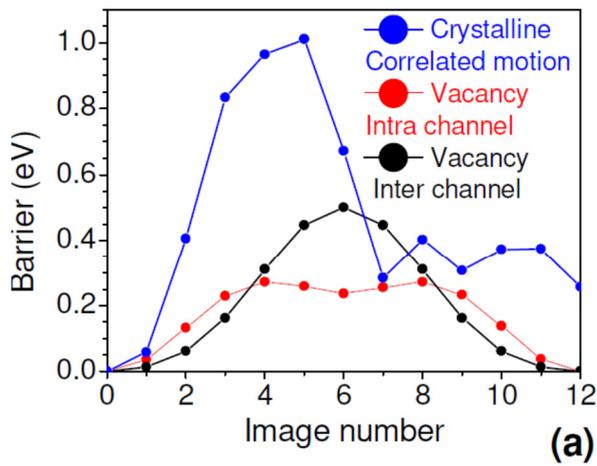
(a)

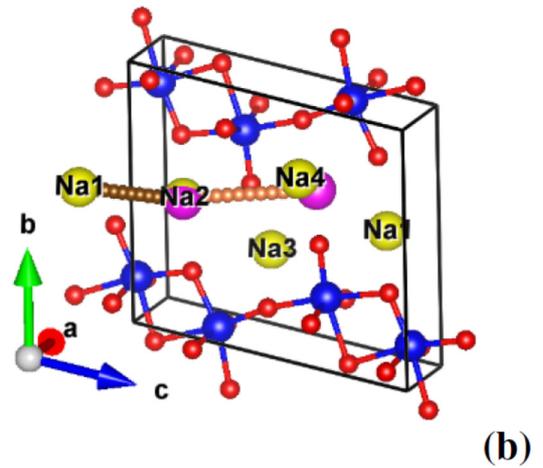
(b)

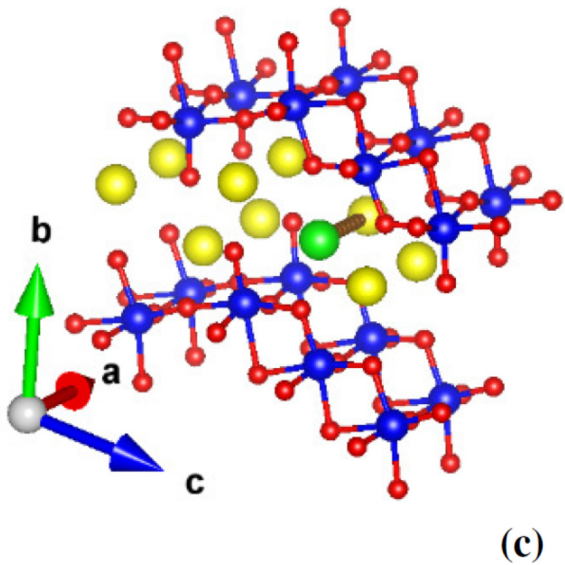
(c)

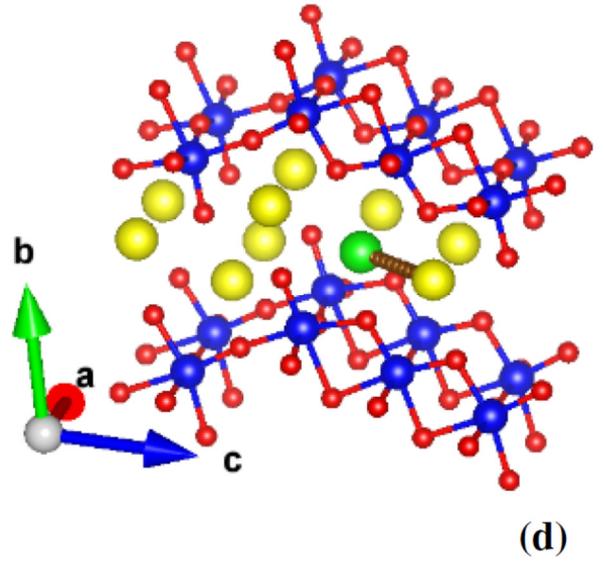
(d)

FIG 7 (Color online) The bond valence energy landscape (BVEL) map of $Na_2Ti_3O_7$ as estimated from the structural data[33] at 300 K. The yellow spheres show the equilibrium positions of Na. The map is shown in the a-c plane, where a-axis is along the one dimensional channel formed by Na atoms.

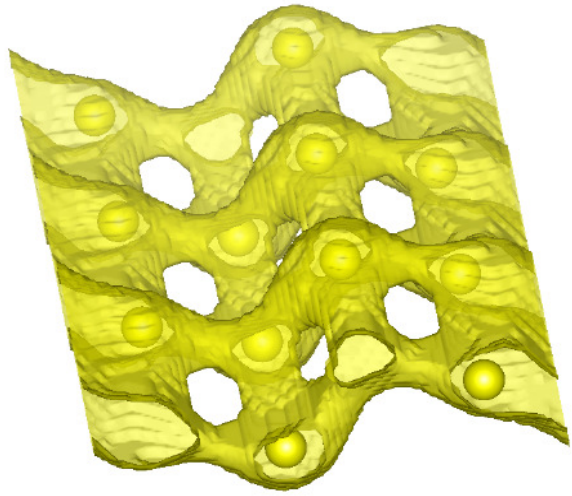

FIG 8 (Color online) The experimentally measured neutron-weighted phonon density of states in $Na_2Ti_3O_7$ at various temperatures. The multi-phonon contribution has been corrected using standard procedure at ISIS.

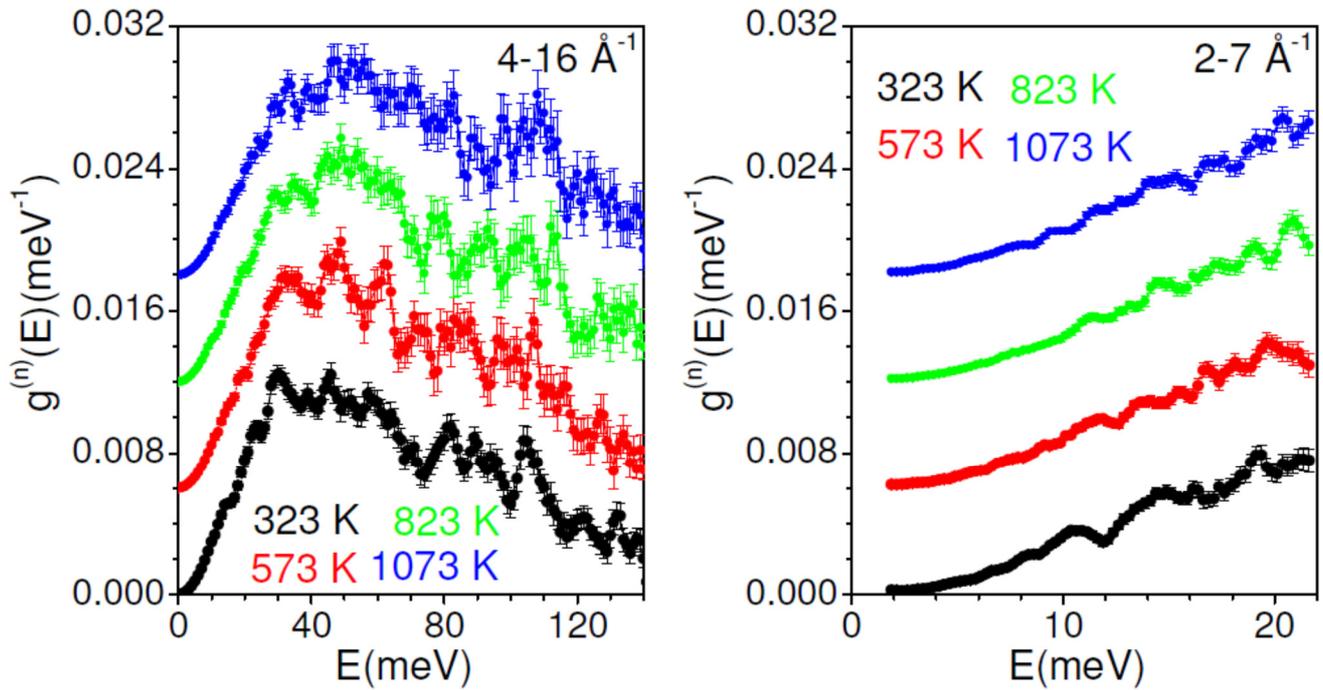

FIG 9 (Color online) (a) The calculated (0 K) and experimental (323 K) neutron-weighted phonon density of states in $Na_2Ti_3O_7$. The phonon spectra have been calculated from ab-initio lattice dynamics. The calculated neutron-weighted partial contributions from various atoms are also shown. Average Q value of Q=10 Å$^{-1}$ has been taken for the Debye-Waller factor. The calculated spectra have been broadened with Gaussian of full width at half maximum of 4 % of incident energy of 180 meV, which correspond to the resolution of the experiment. **(b)** The calculated neutron-weighted phonon density of states of $Na_2Ti_3O_7$ including the effect of the Debye-Waller factor. The calculated neutron-weighted partial contributions from various atoms are also shown. The calculated spectra have been broadened with Gaussian of full width of half maximum of 1 meV. Average Q value of Q=5 Å$^{-1}$ has been taken for the Debye-Waller factor. (c) The calculated total and the partial phonon densities of states of various atoms in $Na_2Ti_3O_7$ from ab-initio lattice dynamics.

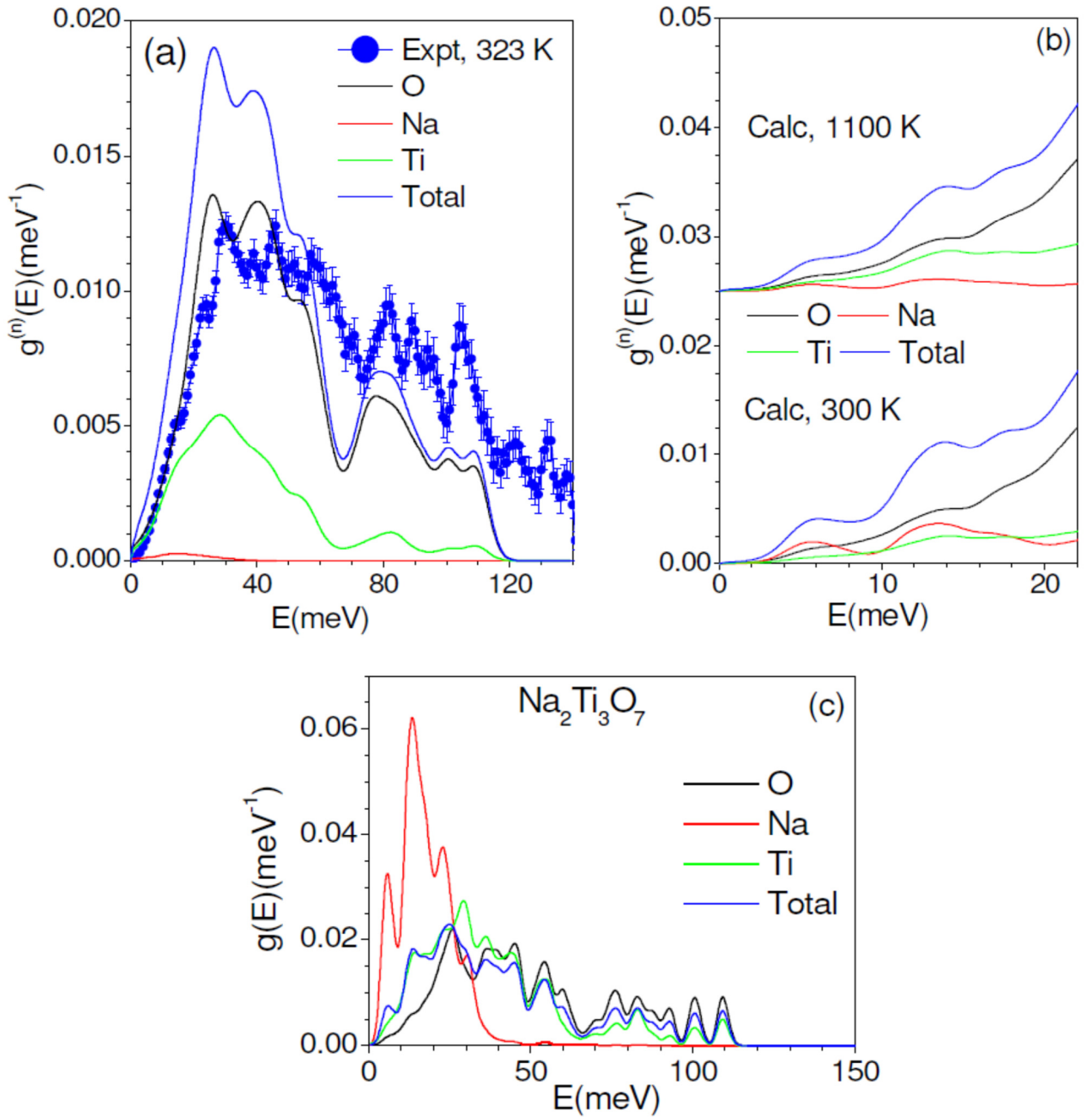

FIG 10 (Color online) (a) The calculated mode Grüneisen parameters Γ of phonons of energy E, averaged over the entire Brillouin zone, as a function of phonon energy (E). Inset in (a) shows the calculated volume thermal expansion coefficients $\alpha_V$. (b) The calculated and experimental (closed symbols) [34] volume thermal expansion in $Na_2Ti_3O_7$. The contribution of phonon modes of energy E, averaged over the entire Brillouin zone, (c) to volume thermal expansion coefficients at 300 K, and (d) to the mean-squared amplitude of various atoms in $Na_2Ti_3O_7$ at 300 K.

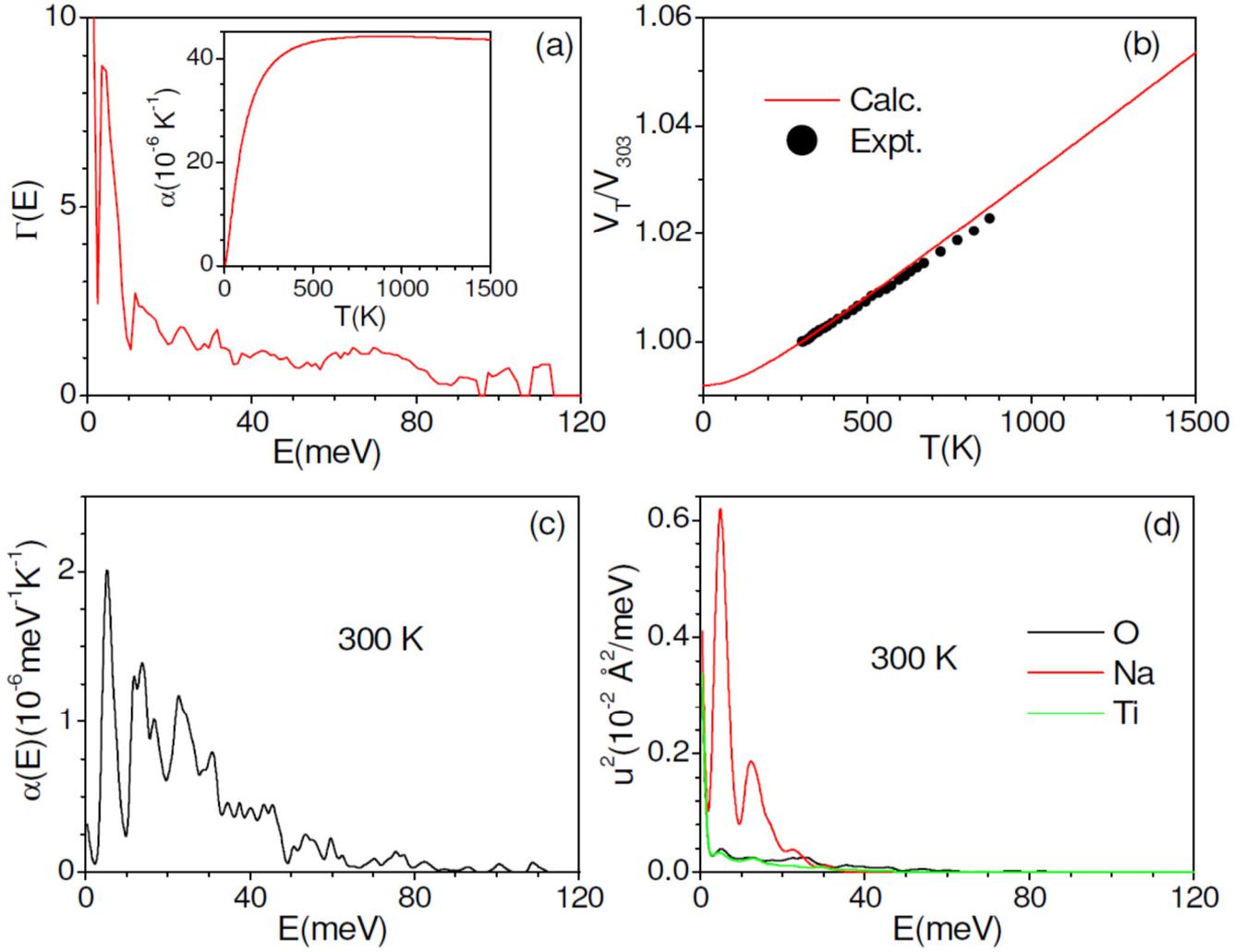